\documentclass[]{emulateapj}

\usepackage{longtable}
\usepackage{graphicx}
\usepackage{amssymb}

\shorttitle{PSN~J09132750+7627410}
\shortauthors{Tartaglia et al.}

\newcommand{\msun}{\,M$_{\odot}$}

\newcommand{\lsun}{\,L$_{\odot}$}
\newcommand{\kms}{\,km~s$^{-1}$}

\newcommand{\ang}{\,$\rm{\AA}$}
\newcommand{\ergs}{\,erg\,s$^{-1}$}
\newcommand{\halpha}{\,H$\alpha$}
\newcommand{\hbeta}{\,H$\beta$}
\newcommand{\hgamma}{\,H$\gamma$}

\newcommand{\psn}{\,PSNJ09+76}

\begin{document}

\title{The supernova impostor PSN~J09132750+7627410 and its progenitor}
\author{
L.~Tartaglia\altaffilmark{1,2},
N.~Elias-Rosa\altaffilmark{1},
A.~Pastorello\altaffilmark{1},
S.~Benetti\altaffilmark{1},
S.~Taubenberger\altaffilmark{3,4},
E.~Cappellaro\altaffilmark{1},
G.~Cortini\altaffilmark{5},
V.~Granata\altaffilmark{1,2},
E.~E.~O.~Ishida\altaffilmark{4,6},
A.~Morales-Garoffolo\altaffilmark{7},
U.~M.~Noebauer\altaffilmark{4},
P.~Ochner\altaffilmark{1},
L.~Tomasella\altaffilmark{1},
S.~Zaggia\altaffilmark{1},
}

\altaffiltext{1}{INAF - Osservatorio Astronomico di Padova, Vicolo dell'Osservatorio 5, 35122 Padova, Italy}
\altaffiltext{2}{Universit\`a degli Studi di Padova, Dipartimento di Fisica e Astronomia, Vicolo dell'Osservatorio 2, 35122 Padova, Italy}
\altaffiltext{3}{European Southern Observatories, Karl-Schwarzschild-Str., D-85748 Garching, Germany}
\altaffiltext{4}{Max-Planck-Institut f\"ur Astrophysik, Karl-Schwarzschild-Str. 1, D-85748 Garching, Germany}
\altaffiltext{5}{Osservatorio Astronomico di Monte Maggiore - Predappio, Italy}
\altaffiltext{6}{Clermont Universit\'e, Universit\'e Blaise Pascal, CNRS/IN2P3, Laboratoire de Physique Corpusculaire, BP 10448, F-63000, Clermont-Ferrand, France}
\altaffiltext{7}{Institut de Ci\`encies de l'Espai (CSIC - IEEC), Campus UAB, Carrer de Can Magrans S/N, 08193 Cerdanyola del Vall\'es, Barcelona, Spain}

\begin{abstract}
We report the results of our follow-up campaign of the supernova impostor PSN~J09132750+7627410, based on optical data covering $\sim250$~d.
From the beginning, the transient shows prominent narrow Balmer lines with P-Cygni profiles, with a blue-shifted absorption component becoming more prominent with time.
Along the $\sim3$~months of the spectroscopic monitoring, broad components are never detected in the hydrogen lines, suggesting that these features are produced in slowly expanding material.
The transient reaches an absolute magnitude $M_r=-13.60\pm0.19\,\rm{mag}$ at maximum, a typical luminosity for supernova impostors.
Amateur astronomers provided $\sim4$~years of archival observations of the host galaxy, NGC~2748.
The detection of the quiescent progenitor star in archival images obtained with the Hubble Space Telescope suggests it to be an $18-20$\msun~white-yellow supergiant.
\end{abstract}


\section{Introduction} \label{intro}
Energetic outbursts of massive stars are often labelled as `supernova impostors' \citep{2000PASP..112.1532V}, since they usually show observational features resembling those of Type IIn supernovae \citep[SNe;][]{1990MNRAS.244..269S,1997ARA&A..35..309F}.
However, SN impostors are non-terminal outbursts of massive stars, extra-galactic counterparts of the `Great Eruption' (occurred in the 19th century) of the Galactic luminous blue variable (LBV) $\eta$-Car.
The mechanisms triggering these events are not yet fully understood \citep{1990A&A...237..409P}. \\

In most cases, SN impostors have been related to major outbursts of LBV stars \citep[e.g.][]{2010AJ....139.1451S,2015MNRAS.447..117T}, since narrow hydrogen lines present in their spectra have inferred expansion velocities comparable to those of LBV winds ($10^{2-3}$~\kms).
LBVs are evolved, luminous (a few $10^{5-6}$\lsun) massive stars ($>30$\msun), very close to the Eddington limit, characterised by large instabilities in the outer layers and high mass losses.
Nonetheless, LBV-like outbursts have also been linked to lower mass stars \citep[see e.g. the case of SN~2008S and other similar transients;][]{2009MNRAS.398.1041B,2009ApJ...705.1364T,2009ApJ...699.1850B,2009ApJ...695L.154B}, while also binary interactions can play an important role in triggering violent mass loss episodes. \\

In this context, we report the results of our follow-up campaign of the SN impostor PSN~J09132750+7627410 (hereafter \psn).
Its discovery was announced on 2015 February 10 through an IAU Central Bureau Astronomical Telegram\footnote{\url{http://www.cbat.eps.harvard.edu/unconf/followups/\\J09132750+7627410.html}} in NGC~2748, which previously hosted other two SNe: the Type Ia SN~1985A \citep{1985IAUC.4031....3W,1987AJ.....93..287W} and the Type Ic SN~2013ff \citep{2013CBET.3647....1B}.
{\psn~was classified} as a SN impostor by \citet{2015ATel.7051....1T} on 2015 February 12. \\

Hereafter, we will adopt a luminosity distance of $D_L=23.8\pm2.0\,\rm{Mpc}$, hence a distance modulus of $\mu=31.88\pm0.18\,\rm{mag}$, as reported by the Extragalactic Distance Database\footnote{\url{http://edd.ifa.hawaii.edu/}} \citep{2009AJ....138..323T}.
For the foreground Galactic extinction we will assume $A(V)=0.073\,\rm{mag}$ \citep{2011ApJ...737..103S}.
We adopted no host galaxy contribution to the total extinction, since in the spectra of \psn~there is no evidence of the narrow \ion{Na}{1D} absorption feature at the recessional velocity of NGC~2748. \\

\section{Observations and data reduction} \label{obsData}
The follow-up campaign of \psn~was carried out using the $1.82\,\rm{m}$ Copernico telescope equipped with AFOSC located at Mount Ekar, Asiago, Italy, the $10.4\,\rm{m}$ Gran Telescopio Canarias (GTC) with OSIRIS and the $2.56\,\rm{~m}$ Nordic Optical Telescope (NOT) with ALFOSC, both located at the Observatorio del Roque de los Muchachos, La Palma, Spain.
Additional photometric data, mostly covering the pre-eruptive phases, were provided by public archives and amateur astronomers. 
The transient was also observed by the UV/Optical Telescope (UVOT) on board of the \textsl{Swift} Gamma-ray Observatory\footnote{\url{http://swift.gsfc.nasa.gov/}}.
UVOT data ({\it uvw2, uvm2, uvw1, U, B, V} bands) were processed using the HEASARC software (HEASoft\footnote{\url{http://heasarc.nasa.gov/lheasoft/}}, version 6.17) following the prescription of \citet{2008MNRAS.383..627P}.
The transient field was also observed by the {\it Hubble Space Telescope} (HST) in the optical domain and the {\it Spitzer Space Telescope} (SST) in Near-Infrared (NIR) and we used these images to perform our analysis on the progenitor of \psn~(see Section~\ref{progenitor}). \\
\begin{figure*} 
\begin{center}
\includegraphics[width=0.95\linewidth]{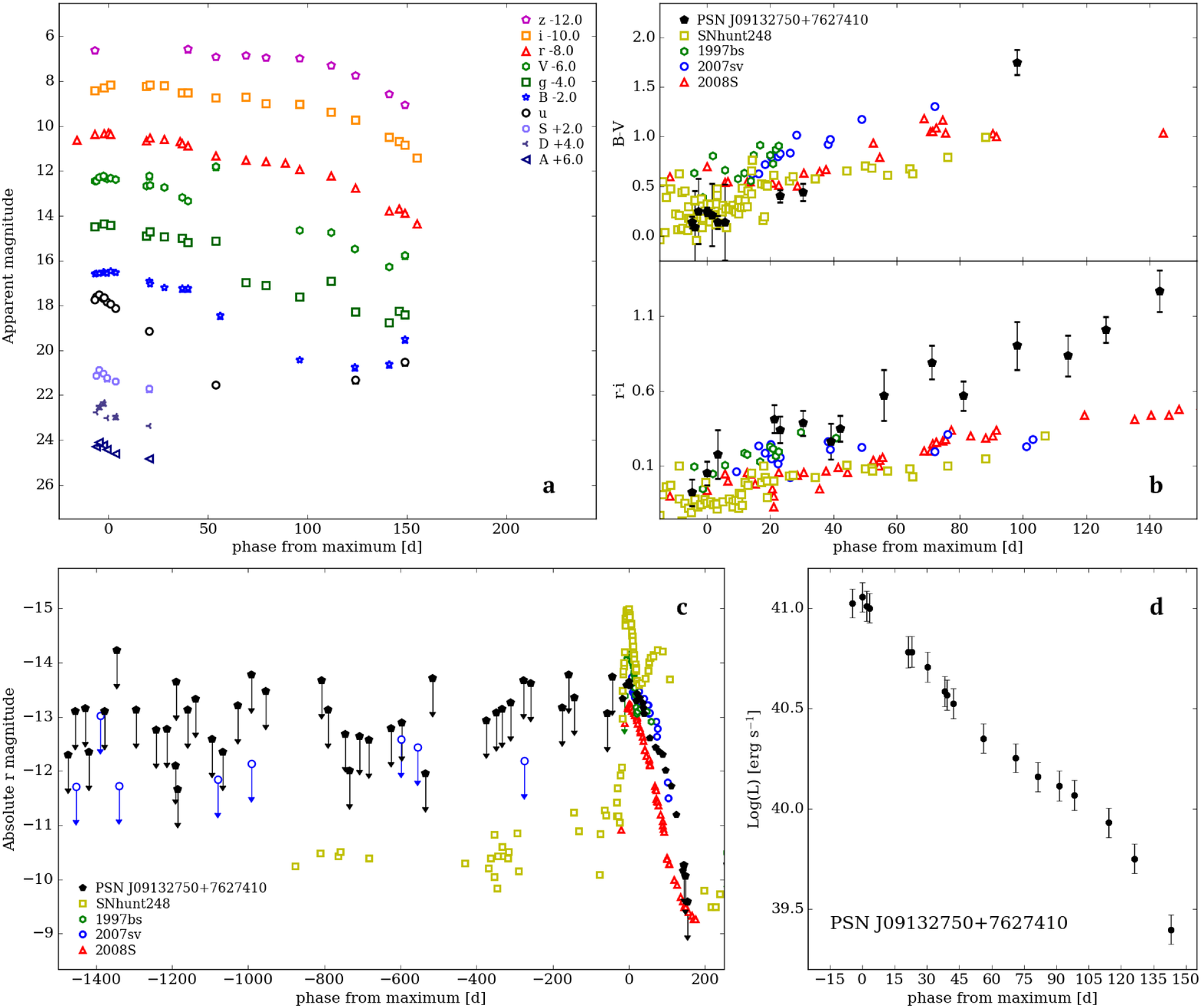}
\caption{{\bf a)} Multi-band light-curves of \psn. Arbitrary constants were added to the different magnitudes. {\bf b)} $B-V$ and $r-i$ colour evolutions of \psn, compared to those of other SN impostors data \citep[2007sv, 1997bs and the peculiar transient 2008S, whose nature is still debated, see e.g.][]{2008ApJ...681L...9P,2009MNRAS.398.1041B,2009ApJ...705L.138P,2009ApJ...697L..49S,2009ApJ...705.1364T,2010MNRAS.403..474W,2011ApJ...741...37K,2012ApJ...750...77S,2015arXiv151107393A}. Distance moduli and reddening estimates of 1997bs ($\mu=31.1\,\rm{mag}$, $A_V=0.093\,\rm{mag}$), 2007sv ($\mu=31.38\,\rm{mag}$, $A_V=0.056\,\rm{mag}$), SNhunt248 ($\mu=31.76\,\rm{mag}$, $A_V=0.140\,\rm{mag}$) and 2008S ($\mu=28.74\,\rm{mag}$, $A_V=1.13\,\rm{mag}$) were taken from \citet{2000PASP..112.1532V}, \citet{2015MNRAS.447..117T}, \citet{2015A&A...581L...4K} and \citet{2009MNRAS.398.1041B} respectively. {\bf c)} Historical absolute {\it r}-band magnitudes of \psn~compared to those of a sample of different transients. {\it r} and {\it i}-band magnitudes for 2007sv, 1997bs, 2008S and SNhunt248 were obtained from the {\it R} and {\it I}-band magnitudes adding the constants 0.16~mag and 0.37~mag respectively, following the transformations given by \citet{2007AJ....133..734B}. {\bf d)} Pseudo-bolometric light-curve of \psn~computed integrating the observed fluxes from the UV to the optical domains. The luminosity is obtained integrating the computed SED with the trapezoidal rule \citep[see Section 3.4 of][for more details]{2013MNRAS.434.1636T} The magnitudes used for this figure are reported in the supplemental online material.} \label{photometry}
\end{center}
\end{figure*}

Ground-based spectroscopic and photometric data were pre-processed in the usual manner (overscan, bias and flat-field corrections) using standard \textsc{iraf}\footnote{\url{http://iraf.noao.edu/}} tasks.
The magnitudes of the transient were measured through a dedicated pipeline \citep[SNOoPY;][]{snoopyref}, and calibrated with reference to a set of field stars.
This local sequence was calibrated with reference to Sloan Digital Sky Survey (SDSS\footnote{\url{http://sdss.org}}) fields obtained during photometric nights, while their $BVRI$ magnitudes were derived from the Sloan magnitudes following the relations in \cite{2008AJ....135..264C}.
Unfiltered data from the amateur astronomers were calibrated to the $r$-band. \\

The spectroscopic calibrations were performed using spectra of standard stars obtained during the same nights (for the flux calibration), and spectra of comparison lamps obtained with the same instrumental set-up used (for the wavelength calibration).
The wavelength calibration was checked using night sky lines, shifting the spectra in case of discrepancy.
The flux calibration was checked using the magnitudes obtained on the closest nights, and a scaling factor was applied when necessary.
Finally, we corrected the spectra of \psn~for the Galactic foreground extinction and for the redshift (z=0.0049) due to the recessional heliocentric velocity of the host galaxy ($1476\pm2$\kms). \\

\section{Photometric analysis}
The outcomes of our photometric measurements are shown in Figure~\ref{photometry}, while the magnitudes in the different bands are reported in the online table. \\

Our {\it i}-band image obtained on 2015 May 25 with ALFOSC (the one with the best seeing among our photometric data, {\it full-width-at-half-maximum}: $\rm{FWHM}\sim0.5$\arcsec) was used to accurately pin-point the position of the transient (RA=09:13:27.408, Dec=$+$76:27:40.91; J2000). \\

About 4~years of archival data were provided mostly by amateur astronomers, and revealed no further eruptive episodes similar to that observed in 2015.
The last non-detection prior to the 2015 event is dated 2015 January 6, $\sim1$~month before our first {\it r}-band point.
The {\it r}-band maximum occurred on 2015 February 16 ($\rm{MJD}=57069.76$), and we consider this as the reference epoch for the phases. \\

After maximum, we note a relatively fast decline in the {\it u}-band light-curve with a slope of 6.8~mag/100~d. 
The light-curves decline more slowly in the other bands, with $4.3\,\rm{mag}/100\,\rm{d}$, $2.8\,\rm{mag}/100\,\rm{d}$ and $3.3\,\rm{mag}/100\,\rm{d}$ in the {\it B}, {\it V} and {\it g}-band, respectively.
The {\it r}-band light-curve shows a decline rate of $1.9\,\rm{mag}/100\,\rm{d}$ in the first $\sim130\,\rm{d}$, while it steepens at later epochs.
A similar behavior is observed in the {\it i} and {\it z}-band light-curves, with $1.3\,\rm{mag}/100\,\rm{d}$ and $0.4\,\rm{mag}/100\,\rm{d}$ in the first $\sim100\,\rm{d}$, while the slopes increase to $5.2\,\rm{mag}/100\,\rm{d}$ and $4.8\,\rm{mag}/100\,\rm{d}$, respectively, at later phases. \\

The evolution of the $B-V$ and $r-i$ colours is shown in Figure~\ref{photometry}, panel~b).
We note that the colours become progressively redder, consistent with a rapid decline of the temperature of the emitting gas.
The colour evolution at early phases is similar to those of other impostors (Figure~\ref{photometry}, although we note that \psn~is much redder than the other transients at phases later than $\sim50$~d from maximum. \\

The {\it r}-band absolute light-curve is shown in panel~c).
Assuming the distance modulus and the extinction reported in Section~\ref{intro}, we infer an {\it r}-band absolute peak magnitude of $-13.60\pm0.19\,\rm{mag}$, corresponding to a pseudo-bolometric luminosity $\gtrsim10^{41}$\ergs~(panel d).
The comparison with the absolute light-curves of similar transients shows that \psn~has an absolute light-curve similar to those of other SN impostors, with the main outburst having a light-curve resembling those of a core-collapse SN, but fainter peak-magnitude. \\

\section{Spectroscopic analysis} \label{spectroscopic}
\begin{figure*} 
\begin{center}
\includegraphics[width=0.95\linewidth]{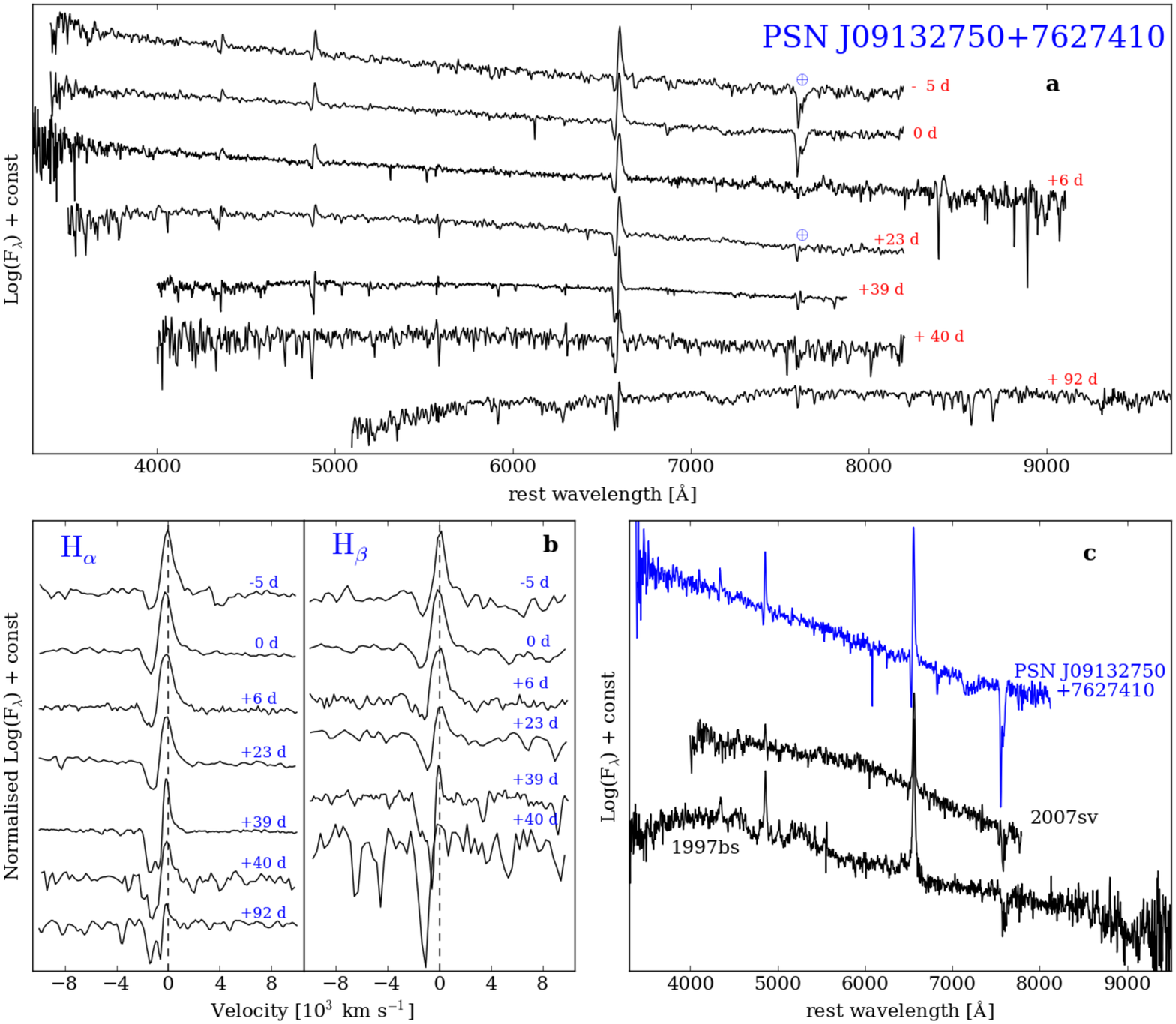}
\caption{{\bf a)} Spectral sequence of \psn. The phases are reported to the right and are referred to the {\it r}-band maximum. $\oplus$ symbols mark the positions of the strongest telluric absorption bands. {\bf b)} Evolution of the \halpha~({\bf left}) and \hbeta~({\bf right}) line profiles in velocity space. Vertical dashed lines mark the rest-wavelength position of the two lines. {\bf c)} Comparison between the spectra of \psn~obtained at the epoch of the {\it r}-band maximum and those of the SN impostors 2007sv and 1997bs at similar phases. All spectra have been dereddened and redshift corrected, and shifted by an arbitrary constant.} \label{spectroscopy}
\end{center}
\end{figure*}

The results of the spectroscopic analysis are shown in Figure~\ref{spectroscopy}.
In panel a) we show the final sequence of spectra obtained during the $\sim3\,\rm{months}$ of our spectroscopic follow-up campaign.
A log of the spectroscopic observations is reported in Table~\ref{specLog}, along with the information on the instrumental set-up used, exposure times and spectral resolutions. \\
\begin{deluxetable*}{@{}ccccccc@{}}
\tablecaption{Log of the spectroscopic observations of \psn. The phases are relative to the maximum.}
\tablehead{\colhead{Date} & \colhead{Phase} & \colhead{Instrumental set-up} & \colhead{Grism or grating} & \colhead{Spectral range} & \colhead{Resolution} & \colhead{Exp. times} \\ 
\colhead{} & \colhead{} & \colhead{} & \colhead{} & \colhead{(\ang)} & \colhead{(\ang)} & \colhead{(s)} }
\startdata
20150212 & -5 & Ekar182$+$AFOSC & 2xgm4 & $3400-8200$ & 14.4 & $2\times1800$ \\
20150216 & 0  & Ekar182$+$AFOSC & gm4 & $3400-8200$ & 14.4 & 2700 \\
20150223 & 6  & NOT$+$ALFOSC & gm4 & $3500-9000$ & 18.1 & 2400 \\
20150311 & 23 & Ekar182$+$AFOSC & gm4 & $3400-8200$ & 14.4 & $2\times2700$ \\
20150326 & 38 & GTC$+$OSIRIS & R1000B & $3600-8000$ &  7.0 & $2\times1800$ \\
20150328 & 40 & Ekar182$+$AFOSC & gm4 & $3400-8200$ & 14.4 & $2\times2700$ \\
20150518 & 92 & GTC$+$OSIRIS & R1000R & $5000-10000$ &  8.0 & $2\times1800$ \\
\enddata
\label{specLog}
\end{deluxetable*}

The spectra at early phases are characterised by a blue continuum with sharp and narrow Balmer lines with P-Cygni profiles, as also noticed by \citet{2015ATel.7172....1H}.
The temperature of the pseudo-continuum, estimated through a black-body fit, rapidly falls from $13400\pm2500\,\rm{K}$ to $3400\pm400\,\rm{K}$, in agreement with the estimated broad-band colour evolution (Figure~\ref{photometry}, panel b). 
The \halpha~profile is characterised by a prominent P-Cygni profile at all phases. \\

In Figure~\ref{spectroscopy} (panel b) we show the detailed evolution of the \halpha~and \hbeta~regions.
The total \halpha~profile is well fitted using a Gaussian component in absorption and a single Lorentzian component in emission.
From this fit, at early phases, we derive a FWHM velocity $<700$\kms~for the emission component and a nearly constant ($900-1000$\kms) expansion velocity, deduced from the positions of the minimum of the absorption components.
Nonetheless, at $+39\,\rm{d}$ and $+92\,\rm{d}$, we note a second absorption component (Figure~\ref{spectroscopy}, panel b), suggesting the presence of a slower shell, moving at $300-450$\kms, which is nearly the same value (within the errors) inferred from the FWHM of the narrow emission component at the same phases. 
The absence of this component at earlier phases, as well as in other spectra at similar phases, is most likely due to insufficient resolution.
The velocity inferred from the blue wing of the faster absorption components in the $+39\,\rm{d}$ spectrum is consistent with that of the red wing of the \halpha~emission profile.
We do not detect the same features in the \hbeta~profile, although this might be related to the lower {\it signal-to-noise ratio} (SNR) of our spectra at the corresponding wavelengths. \\

The \hgamma~line is also identified at early phases.
In our $+39\,\rm{d}$ spectrum, we also identify \ion{Fe}{2} (multiplets 27, 37, 38, 42, 48, 49), \ion{Ba}{2} $\lambda\lambda$4554.0, 4934.1, \ion{Na}{1D} $\lambda\lambda$5889.9, 5895.9, \ion{O}{1} at 7774\ang~and \ion{Ca}{2} (H and K), \ion{Ti}{2} (multiplets 1, 19, 20, 31, 34, 41, 51, 82, 105) and, tentatively, the \ion{Ba}{1} (multiplet 2).
We note that our last ($+92$~d) spectrum has a much redder continuum, and broad features appear, most likely due to TiO molecular bands in the $6100-6400$\ang~and $7000-7400$\ang~regions. \\

\section{The progenitor star of \psn} \label{progenitor}
\begin{figure*} 
\begin{center}
\includegraphics[width=0.71\linewidth]{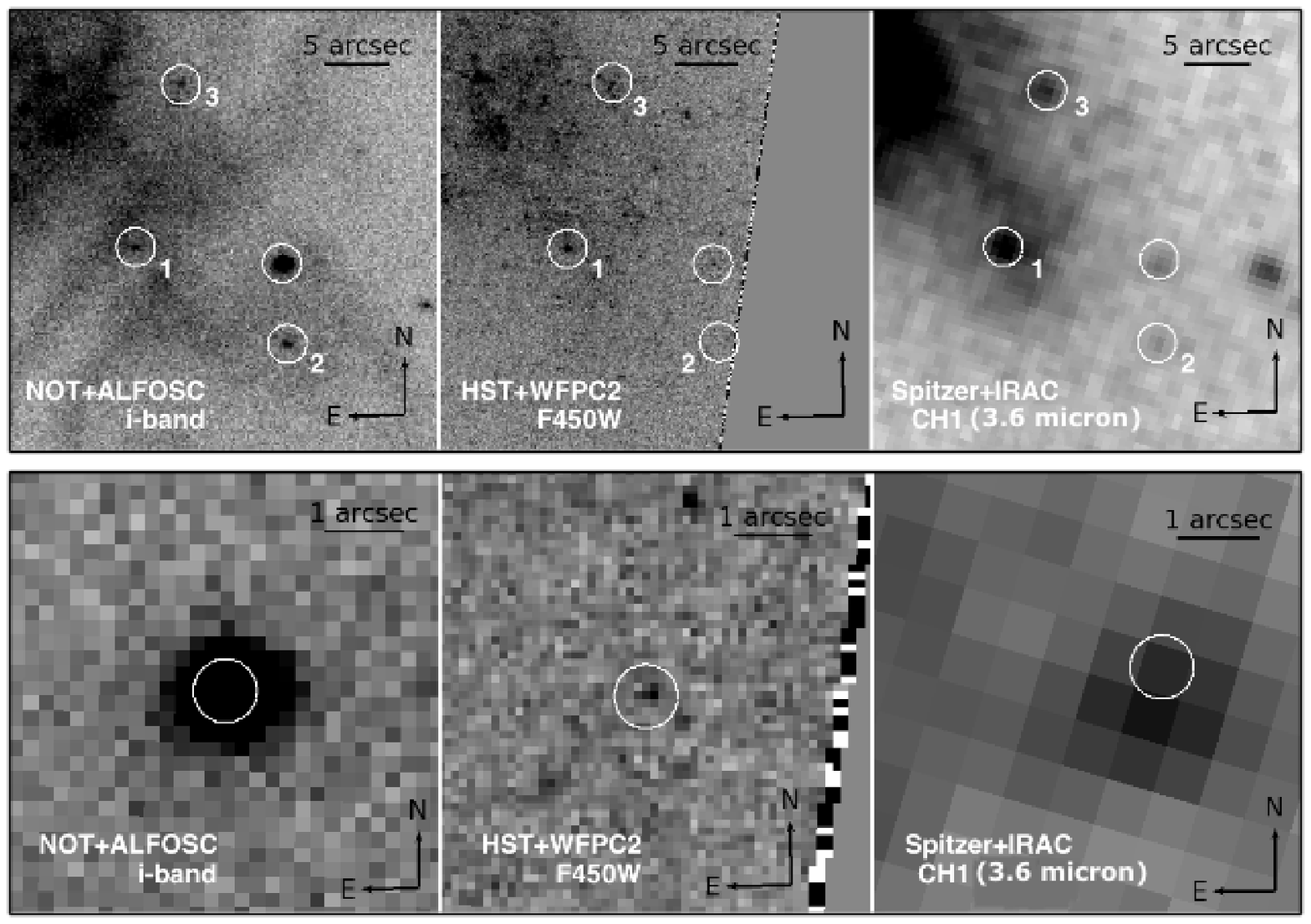}
\caption{Pre-eruption archival images of the site of \psn. {\bf Top:} Details of the ALFOSC {\it i}-band image obtained on 2015 May 25, the HST $F450W$ image and the SST image obtained on 2010 April 12. The positions of three reference stars are also shown. {\bf Bottom:} Zoom in of the position coincident with that of \psn~in the ALFOSC images. The positions in the HST and SST images are those computed using the \textsc{geomap} and \textsc{geoxytran} \textsc{iraf} tasks.} \label{archival_images}
\end{center}
\end{figure*}
Deep archival HST images\footnote{\url{http://archive.stsci.edu/hst/search.php}} of the host galaxy were obtained on 2001 July 06 ({\sl HST} ID proposal: 9042; PI: S.~Smartt) using the Wide Field Planetary Camera 2 (WFPC2) with the $F450W$ and $F814W$ filters.
Also three sets of mid-infrared (MIR) SST images were obtained on 2009 December 2, 2010 April 12 (ID proposal: 61063; PI: K.~Sheth) and 2014 January 2 (ID proposal: 10046; PI: D.~Sanders) using the Infra-Red Array Camera (IRAC) with the $3.6\,\mu\rm{m}$ and $4.5\,\mu\rm{m}$ channels (channels 1 and 2 respectively). These data were retrived from by the {\sl Spitzer} Heritage Archive\footnote{\url{http://sha.ipac.caltech.edu/applications/Spitzer/SHA/}} (SHA), hence fully co-added and calibrated.
Drizzled HST images (resampled to a uniform grid to correct geometric distortions, with $\sim0.1\,\rm{arcsec}\,\rm{pixel^{-1}}$) obtained through the {\sl Hubble} Legacy Archive\footnote{\url{http://hla.stsci.edu/}} (HLA) and {\it Post Basic Calibrated Data} (pbcd) SST images (with $\sim0.6\,\rm{arcsec}\,\rm{pixel^{-1}}$), were aligned to our {\it i}-band ALFOSC image (with 0.19~arcsec~pixel$^{-1}$) obtained on 2015 May 8.
Geometrical alignments were performed using the \textsc{iraf} task \textsc{geomap}, and a maximum of 17 common point-like sources between all sets of images. The errors of the geometric transformations are the root-mean-square (\textsc{rms}) uncertainties given by \textsc{geomap}, obtaining a precision in the corresponding position of \psn~ of $<0.020$\arcsec\footnote{We obtained the following transformation uncertainties: $\rm{(x,y)}=0.013$\arcsec, 0.018\arcsec~(11 stars used) and $\rm{(x,y)}=0.019$\arcsec, 0.020\arcsec~(15 stars used) in the $F450W$ and $F814W$ images respectively, $\rm{(x,y)_1}=0.066$, 0.072\arcsec~(16 stars), $\rm{(x,y)_2}=0.047$, 0.084\arcsec~(16 stars), $\rm{(x,y)_1}=0.078$, 0.060\arcsec~(17 stars), $\rm{(x,y)_2}=0.066$, 0.072\arcsec~(16 stars), and $\rm{(x,y)_1}=0.028$, 0.034\arcsec~(12 stars), $\rm{(x,y)_2}=0.072$, 0.090\arcsec~(15 stars) for the channel 1 and 2 of images obtained with SST on 2009 December 02, 2010 April 12 and 2014 January 2, respectively.}. \\
\begin{figure}
\begin{center}
\includegraphics[width=0.95\linewidth]{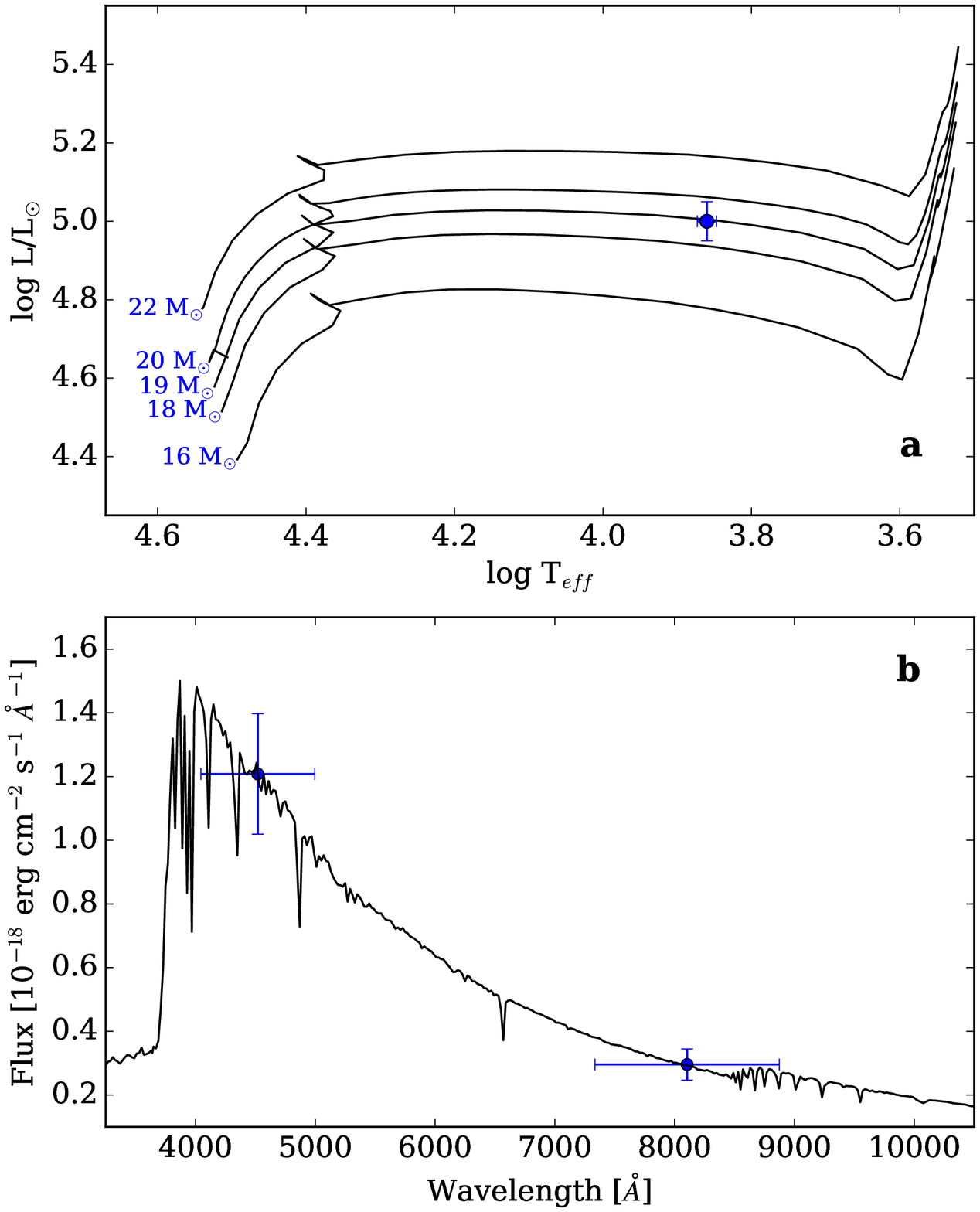}
\caption{{\bf a)} Hertzsprung-Russel diagram showing the position of the \psn~progenitor candidate (blue circle). Evolutionary tracks (from 16~\msun~to 23~\msun) computed using the Cambridge \textsc{stars} \citep{2004MNRAS.353...87E} models are also shown. The tracks were obtained assuming solar metallicity. {\bf b)} Observed SED of the candidate progenitor (blue circles, with uncertainties). An {\textsc ATLAS} synthetic spectrum for a star with $T_{\rm{eff}}=7250\,\rm{K}$, $\log{(g)}=1.5$ and solar metallicity is also plotted.} \label{hdr}
\end{center}
\end{figure}

We identify a source at the position of the transient (within the transformation errors) in both HST and SST images although, given the different spatial resolution, the source in the SST images is most likely the blend of multiple sources (see Figure~\ref{archival_images}).
HST magnitudes and errors of the source in the \textsc{vegamag} flight-system ($F450W=24.27\pm0.17$ and $F814W=24.00\pm0.18\,\rm{mag}$) were obtained using the \textsc{dolphot} package\footnote{\url{http://americano.dolphinsim.com/dolphot/}. We used the WFPC2 Module, v.2.0 (updated in November 2014)} \citep{2000PASP..112.1383D}. 
The \textsc{doplhot} output provides also a set of parameters which can be used to interpret the nature of the different sources. 
The detected source has a `object-type' flag of `1', which means that it is likely stellar. 
Accounting for the extinction ($A_{F450W}=0.093\,\rm{mag}$, $A_{F814W}=0.044\,\rm{mag}$) and distance modulus reported before, we estimate for the progenitor candidate the absolute magnitudes of $-7.71\pm0.25\,\rm{mag}$ and $-7.95\pm0.25\,\rm{mag}$ for filters $F450W$ and $F814W$, respectively. \\

The source magnitudes in the SST images were measured using the \textsc{spot}\footnote{\url{http://ssc.spitzer.caltech.edu/warmmission/propkit/spot/}} package, integrating the flux in a $4\times4$ pixels area around the position of the progenitor, which included the whole extended source seen in Figure~\ref{archival_images} (bottom-right panel). 
We obtained $3.1\times10^{-19}$ and 1$.7\times10^{-19}$, $3.0\times10^{-19}$ and $0.8\times10^{-19}$, $2.7\times10^{-19}$ and $1.4\times10^{-19}\,\rm{erg}\,\rm{s^{-1}}\,\rm{cm^{-2}}\,\ang^{-1}$, resulting in the approximate magnitudes of $18.3\,\rm{mag}$ and $18.0\,\rm{mag}$, $18.3\,\rm{mag}$ and $18.8\,\rm{mag}$, $18.5\,\rm{mag}$ and $18.2\,\rm{mag}$ in the channel 1 and 2 from the images obtained with SST on 2009 December 2, 2010 April 12 and 2014 January 2, respectively.
As we can see, there is no clear evidence of an excess of emission (e.g. due to dust formation) between years 2009 and 2014 at $3.6\,\mu\rm{m}$, but some variability is perceptible at $4.5\,\mu\rm{m}$. 
However, we remark that the resulting magnitudes are likely due to the contribution of multiple sources in the vicinity of the progenitor. \\

\section{Discussion and conclusions} \label{discussion}
In this paper we have discussed the outcomes of our photometric and spectroscopic monitoring of the optical transient \psn.
Its {\it r}-band absolute magnitude at peak ($M_r=-13.60\pm0.19\,\rm{mag}$) is consistent with that expected during giant eruptions of massive stars.
In Figure~\ref{photometry} (panels b and c), we have shown that the light-curves of these transients may differ in terms of absolute magnitudes and colour evolutions. \\

At all phases, the spectra of \psn~are characterised by prominent Balmer lines showing sharp and narrow P-Cygni profiles. 
In our $+39\,\rm{d}$ and $+92\,\rm{d}$ best resolution spectra, we notice two components in absorption in the \halpha~profile.
Multiple absorptions were also reported in a moderate resolution spectrum of \psn~obtained by \citet{2015ATel.7172....1H} around maximum.
The clear detection of two absorption components suggests the presence of two expanding shells moving at different velocities (namely $\sim1000$\kms~and $\sim340-450$\kms).
The velocity inferred for the slower component is comparable with the FWHM velocity of the narrow emission component visible in both $+39\,\rm{d}$ and $+92\,\rm{d}$ spectra.
This shell, initially photo-ionised, quickly recombines, and contributes to a large fraction of the narrow \halpha~flux.
The higher velocity component is likely due to the presence of a further shell, moving at a higher velocity.
The absence (or weakness) of the emission component is indicative that this shell is likely outer. 
Despite the lack of any evidence of shell-shell collision from our data of \psn, we speculate that one of the shells may have been expelled in past outbursts. \\

The analysis of deep pre-outburst archival images obtained with HST and SST revealed the presence of a source at the position of \psn~(within the errors of our relative astrometry).
In order to better constrain the nature of the candidate progenitor, we estimated the oxygen abundance at the position of the source following \citet{2009A&A...508.1259H} and the prescriptions of \citet{2004ApJ...613..898T} and \citet{2004A&A...425..849P} \citep[see also][for more details]{2015MNRAS.449.1954P}, which resulted in $12+\log{[O/H]}\simeq8.59$~dex. 
This value is very close to the solar abundance, $8.69\pm0.05\,\rm{dex}$ \citep{2009ARA&A..47..481A}. 
However, following \cite{2009MNRAS.395.1409S}, who sets $12+\log{[O/H]}=8.4\,\rm{dex}$ as the dividing line between solar and subsolar metallicity evolutionary tracks, we conclude that the environment of \psn~is likely at solar metallicity. \\

The comparison of the observed photometry with {\textsc ATLAS} synthetic spectra\footnote{\url{http://www.stsci.edu/hst/observatory/crds/castelli\_kurucz\\\_atlas.html}} \citep{2004astro.ph..5087C} suggests an effective temperature of the quiescent progenitor of $\sim7250$~K (Figure~\ref{hdr}), resulting in a luminosity of $\rm{Log}(L/$\lsun$)\simeq5$ (with a bolometric correction derived from the synthetic spectrum of $-0.012\,\rm{mag}$).
These values of colour and luminosity of the recovered precursor candidate suggest a progenitor mass in the range $18-20$~\msun~(see Figure~\ref{hdr}). 
Thus, the progenitor candidate of \psn~is likely a moderate-mass white-yellow supergiant, possibly an A-F type star. \\

Similar results were also obtained by \citet{2015A&A...581L...4K} and \citet{2010AJ....139.1451S}, who found moderate-mass progenitors ($M\lesssim20$\msun) for SNhunt248 and UGC~2773-OT, respectively.
The implication from these observational constraints is that not only extremely massive stars can experience large outbursts and hence produce SN impostors.

\section*{acknowledgments}
We thank G.~Cortini (Monte Maggiore Observatory - Predappio - ITALY), F.~Martinelli and R.~Mancini for their observations obtained on 2011 April 2 and 2013 September 2 respectively. ST and UMN acknowledge support by TRR33 `The Dark Universe' of the German Research Foundation (DFG). LT, NER, AP and SB are partially supported by the PRIN-INAF 2014 (project `Transient Universe: unveiling new types of stellar explosions with PESSTO'). Based on observations made with: The Cima Ekar 1.82~m Telescopio Copernico of the Istituto Nazionale di Astrofisica of Padova, Italy. The Gran Telescopio Canarias (GTC) operated on the island of La Palma at the Spanish Observatorio del Roque de los Muchachos of the Instituto de Astrofisica de Canarias. The Nordic Optical Telescope (NOT), operated by the NOT Scientific Association at the Spanish Observatorio del Roque de los Muchachos of the Instituto de Astrofisica de Canarias. Based in part on data collected at Okayama Astrophysical Observatory and obtained from the Subaru Mitaka Okayama Kiso Archive (SMOKA), operated by the Astronomy Data Center, National Astronomical Observatory of Japan. Based in part on observations made with the NASA/ESA Hubble Space Telescope, obtained from the Hubble Legacy Archive, which is a collaboration between the Space Telescope Science Institute (STScI/NASA), the Space Telescope European Coordinating Facility (ST-ECF/ESA) and the Canadian Astronomy Data Centre (CADC/NRC/CSA). This work is based in part on observations made with the Spitzer Space Telescope, operated by the Jet Propulsion Laboratory, California Institute of Technology under a contract with NASA. The Swift analysis software is part of HEASoft (the High Energy Astrophysics Software), which encompasses FTools and XANADU (i.e., XSPEC, XRONOS and XIMAGE for spectral, timing and image analysis respectively). \textsc{iraf} is distributed by the National Optical Astronomy Observatory, which is operated by the Associated Universities for Research in Astronomy, Inc., under cooperative agreement with the National Science Foundation. \textsc{dolphot} is a stellar photometry package adapted from \textsc{HSTphot} for general use.

{}

\end{document}